\documentclass[aps, prd, twocolumn,a4paper]{revtex4}
\usepackage{ulem}
\usepackage{color}
\usepackage{float}
\usepackage{graphicx}
\usepackage{epsfig}
\usepackage{epstopdf}
\usepackage{amsmath}
\usepackage{subfig}
\usepackage[dvipsnames]{xcolor}
\newcommand{\be}{\begin{equation}}
\newcommand{\ee}{\end{equation}}
\newcommand{\ba}{\begin{eqnarray}}
\newcommand{\ea}{\end{eqnarray}}
\newcommand{\nn}{\nonumber\\}
\usepackage{graphicx}
\usepackage{color}
\begin{document}
\title{Relativistic hydrodynamics with momentum dependent relaxation time}
\author{Sukanya Mitra}
\affiliation{Department of Nuclear and Atomic Physics, Tata Institute of Fundamental Research, Homi Bhabha Road, Mumbai 400005, India}

\begin{abstract}

A second order relativistic hydrodynamic theory has been derived using momentum dependent relaxation time in the relativistic transport equation.
In order to do that, an iterative technique of gradient expansion approach, namely the Chapman-Enskog (CE) expansion of the particle distribution function 
has been employed. The key findings of this work are, (i) momentum dependent relaxation time in collision term results in an extended Landau
matching condition for the thermodynamic variables,
(ii) the result from numerical solution of Boltzmann equation lies somewhere in between the 
two popular extreme limits : linear and quadratic ansatz, indicating a fractional power of momentum dependence in relaxation time to be appropriate,
(ii) an equivalence has been established between the iterative gradient expansion method like
CE and the well known moment approach like Grad's 14-moment method. 

\end{abstract}
\maketitle

\section{Introduction}

In the last few decades of exploring the deconfined quark-gluon plasma (QGP) at heavy-ion experimental facilities like RHIC and LHC, relativistic dissipative 
hydrodynamics has offered itself to be a trusted theory that considerably explains the experimental observables \cite{Landau,Blaizot:1986bh,Bjorken:1982qr,Weinberg:1971mx,Jeon:2015dfa,Song:2007ux}. 
It can be best outlined as 
an effective long-wavelength theory which describes the dynamics of conserved, macroscopic quantities. However, in 
order to derive a formalism of this macroscopic theory, a microscopic theory is essential to begin with that captures the dynamical interaction of the system.
Boltzmann transport equation is serving the purpose for quite some time which describes the evolution of the single particle distribution function
by a collision term that essentially includes the microscopic interactions of the system. The hydrodynamic equations so obtained provides the space-time evolution
of thermodynamic and dissipative quantities, in which microscopic interactions enter through the transport coefficients.

However, there are several existing methods of extracting hydrodynamic equations from transport theory. The two popularly
used competing methods are, (i) an iterative technique of successive gradient expansion (order by order) of the out of equilibrium distribution function -
Chapman-Enskog (CE) method and (ii) taking moment integral directly from the transport equation - Grad's 14-moment method. 
Analysis from both the methods along with entropy maximization technique present a vast deal of work in the existing literature \cite{Muronga:2001zk,Denicol:2010xn,Denicol:2012cn,Jaiswal:2013vta}. 
Clearly, a correspondence between these different approaches is certainly desirable in order to have a unique theory.

The difficulty in solving the transport equation comes from the non-linearity of the collision term (it includes product of distribution functions).
Different approximations are made to linearize the theory among which the relaxation time approximation of particle distribution function proposed in \cite{Relax}, 
is one of the simple yet efficient methods in near equilibrium situations. However, \cite{Relax} as well as most of the works concerning relaxation time
approach consider it to be independent of particle momenta. This assumption has two serious drawbacks. First it does not consider the microscopic
momentum anisotropies (i.e, assumes that equilibrium restoration times of phase space distribution functions belonging to all particle momenta are the same),
and secondly, it results in identical relaxation times for microscopic particle distributions and macroscopic fields like viscous flow, while the later is expected to have a slower 
relaxation rate with respect to the former depending on the length and time scales of the concerning systems. There are a few attempts of formulating
wave-number dependent viscous coefficients \cite{Hansen}, but it lacks a fully developed hydrodynamic theory that includes momentum dependent microscopic dynamics.

In this present work a second order relativistic hydrodynamic theory has been developed including momentum dependent relaxation time ($\tau_R$) of particle distribution function
for the first time, using CE method for a conformal system with no conserved charges.
The out of equilibrium distribution function derived in this manner shows serious consequences on the Landau matching condition of macroscopic variables.
An interesting finding of this work
turns out to be the possibility of fractional power of momentum dependence of $\tau_R$ between two accepted extreme limits - linear and quadratic ansatz, as predicted by
\cite{Dusling:2009df}.
Finally, the derived hydrodynamic equation has been compared with the same obtained from moment method and an equivalence between the two approaches has been attempted
thereafter. 

The manuscript is organized as follows. Section II deals with the derivation of momentum dependent second order relativistic viscous hydrodynamics, the extended
Landau matching condition and the equivalence between the CE and Grad's 14 moment method respectively in three different subsections. Section III provides a quantitative
estimation of the effect of momentum dependent relaxation time on pressure anisotropy of the system. Finally, in section IV the article has been summarized by 
a conclusion and possible outlook.

\section{Formalism}
\subsection{Second order relativistic hydrodynamics with momentum dependent relaxation time}
To construct the necessary formalism, I begin with the relativistic transport equation in the following form,
\be
p^{\mu}\partial_{\mu}f(x,p)=C[f]=-\frac{p^{\mu}u_{\mu}}{\tau_R}\delta f~.
\label{RTE}
\ee
Here $f$ is the single particle momentum distribution which is function of particle 4-momenta $p^{\mu}$ and space-time variable $x$. $C[f]$ is the 
collision term expressed in terms of relaxation time $\tau_R$ and the out of equilibrium part of the distribution function
$\delta f=f^{(0)}\phi$, with $f^{(0)}$ being the equilibrium distribution function and $\phi$ being the deviation from it.

The momentum dependence of $\tau_R$ is expressed as a power law of the scaled particle energy $\tau_p=\frac{p^{\mu}u_{\mu}}{T}$ much in the same
line as indicated in \cite{Denicol:2014vaa,Jaiswal:2016sfw},

\be
\tau_R(x,p)=\tau_R^0(x) \tau_p^n~,
\label{relax}
\ee

where $\tau_R^0$ is the momentum independent part of relaxation time and $n$ is  a parameter specifying the power of scaled energy. 
In \cite{Denicol:2014vaa} $\tau_R^0$ is identified as a time scale proportional to the mean-free path of the system which is typically the microscopic time scale. 
$u^{\mu}$ and $T$ are respectively
the hydrodynamic four velocity and temperature of the system. $n=0$ and $n=1$ cases
are termed as linear and quadratic ansatz with further explanations to follow. 

Next, we proceed with the well known CE method of obtaining the unknown
particle distribution function of $r^{th}$ order using its known $(r-1)^{th}$ order values in an iterative method \cite{Degroot}.
Expanding the distribution function with the help of a parameter (typically, the Knudsen number which is a ratio between the mean free path and the macroscopic length scale
of the system) and comparing equal powers of this parameter from Eq.(\ref{RTE}) we have,
\be
\tau_p(Df)^{(r)}+\Pi^{\mu}\nabla_{\mu}f^{(r-1)}=-\frac{\tau_p}{\tau_R}f^{(0)}\phi^{(r)}~,~~r\geq1
\label{CE}
\ee
where,
\ba
&&(Df)^{(r)}=\sum_{s=1}^r \bigg[\frac{\partial f^{(r-s)}}{\partial T}(DT)^{(s)}+\frac{\partial f^{(r-s)}}{\partial u^{\mu}}(Du^\mu)^{(s)}\bigg]\nn
&&~~~~~~+\sum_{s=2}^r \bigg[\frac{\partial f^{(r-s)}}{\partial(\nabla^{\mu}T)}\big\{D(\nabla^{\mu}T)\big\}^{(s)}+\cdots\bigg]+\cdots~.
\ea
Here $f^{(r)}=f^{(0)}\phi^{(r)}$ is the $r^{th}$ order gradient correction to $f$. 
$\Pi^{\mu}=\frac{p^{\mu}}{T}$ is the scaled particle four-momenta. $D=u^{\mu}\partial_{\mu}$ and $\nabla^{\mu}=\Delta^{\mu\nu}\partial_{\nu}$ are temporal and spatial 
counterparts of the total space-time
derivative $\partial^{\mu}=u^{\mu}D+\nabla^{\mu}$, defined with the projection operator $\Delta^{\mu\nu}=g^{\mu\nu}-u^{\mu}u^{\nu}$. Throughout the analysis the metric of
the system has taken to be $g^{\mu\nu}=(1,-1,-1,-1)$.

In order to solve Eq.(\ref{CE}), we need to define the thermodynamic identities of the system that follow from conservation of energy momentum tensor $T^{\mu\nu}$.
With vanishing bulk viscosity and conserved currents it is expressed as,

\be
T^{\mu\nu}(x)=g\int d\Gamma_p p^{\mu}p^{\nu}f=\big\{\epsilon u^{\mu}u^{\nu}-P\Delta^{\mu\nu}\big\}+\pi^{\mu\nu}~,
\label{enmom}
\ee
with $d\Gamma_p=\frac{d^3p}{(2\pi)^3p^0}$ and $g$ respectively as the phase space factor and degeneracy of the system. $\epsilon=u_{\mu}u_{\nu}T^{\mu\nu}$, $P=-\frac{1}{3}\Delta_{\mu\nu}T^{\mu\nu}$ 
and $\pi^{\mu\nu}=\Delta^{\mu\nu}_{\alpha\beta}T^{\alpha\beta}$ are the energy density, pressure and shear stress tensor respectively with the traceless projection operator
$\Delta^{\mu\nu}_{\alpha\beta}=\frac{1}{2}\{\Delta^{\mu}_{\alpha}\Delta^{\nu}_{\beta}+\Delta^{\nu}_{\alpha}\Delta^{\mu}_{\beta}-\frac{2}{3}\Delta^{\mu\nu}\Delta_{\alpha\beta}\}$. 
The first two terms in the last expression of (\ref{enmom}) constitute the equilibrium part $T^{\mu\nu}_0$ and the remaining last term denotes dissipative part $\Delta T^{\mu\nu}$. 
The conservation equation $\partial^{\nu}T_{\mu\nu}=0$ contracted 
with $u^{\mu}$ and $\Delta^{\alpha\mu}$ respectively, gives the equation of energy density and velocity of the system as follows,

\ba
&&D\epsilon=-(\epsilon+P)\partial_{\mu}u^{\mu}+\pi^{\mu\nu}\sigma_{\mu\nu} ~,
\label{iden-1}\\
&&Du^{\mu}=\frac{1}{(\epsilon+P)}\big\{\nabla^{\mu}P-\Delta^{\mu}_{\nu}\nabla_{\rho}\pi^{\nu\rho}+\pi^{\mu\nu}Du_{\nu}\big\}~,
\label{iden-2}
\ea
with $\sigma_{\mu\nu}=\nabla_{\langle{\mu}}u_{\nu\rangle}$. The notation $\langle\rangle$ denotes the traceless irreducible tensors of rank-1 and 2 defined as $A_{\langle\mu\rangle}=\Delta_{\mu\nu}A^{\nu}$ 
and $ A_{\langle\mu} B_{\nu\rangle}  =\Delta_{\mu\nu\alpha\beta}A^{\alpha}B^{\beta}$ respectively.

Putting $r=1$ in Eq.(\ref{CE}), employing Eq.(\ref{relax}) and using identities (\ref{iden-1}) and (\ref{iden-2}), the first order correction to the particle distribution
function is obtained in terms of velocity gradient,

\be
\phi^{(1)}=\tau_R^0\tau_p^{n-1}\Pi^{\langle\mu}\Pi^{\nu\rangle}\sigma_{\mu\nu}~.
\label{phi1}
\ee
Clearly one can see, this first order correction is a linear and a quadratic function of particle momenta for $n=0$ and $n=1$ respectively and hence
bears the name. Putting Eq.(\ref{phi1}) in the well known first order contribution of shear viscous stress tensor,

\be
\pi^{(1)\mu\nu}=g\int d\Gamma_p p^{\langle\mu}p^{\nu\rangle}f^{(0)}\phi^{(1)}=2\eta\sigma^{\mu\nu}~,
\label{shearstress1}
\ee
provides the following relation that constrains the momentum independent part $\tau_R^0$ by the shear viscosity $\eta$ scaled over entropy density $s=\frac{(\epsilon+P)}{T}$
of the system,

\be
\tau_R^0=\frac{\eta/s}{T}\frac{5!}{(n+4)!}~.
\label{shear1}
\ee
With $r=2$ in Eq.(\ref{CE}) and doing a bit of algebra, the second order correction to the particle distribution function is obtained as the following,

\ba
&&\phi^{(2)}=\tau_R^0\bigg[-\frac{\tau_p^{n+1}}{12P}\pi^{(1)\mu\nu}\sigma_{\mu\nu}-\frac{\tau_p^n}{4P}\Pi_{\langle\nu\rangle}\nabla_{\mu}\pi^{(1)\mu\nu}\nonumber\\
&&+\frac{5!}{(n+4)!}\frac{1}{8P}\bigg\{-\tau_p^{2n-1}\Pi^{\mu}\Pi^{\nu}D\pi^{(1)}_{\mu\nu}\nn
&&-2\tau^{2n-1}_p\Pi^{\mu}\Pi^{\nu}\pi^{(1)}_{\mu\nu}(\partial\cdot u)-\tau^{2n-2}_p\Pi^{\alpha}\Pi^{\beta}\Pi^{\mu}\nabla_{\mu}\pi^{(1)}_{\alpha\beta}\nn
&&-\big((n-1)\tau_p^{2n-3}-\tau_p^{2n-2}\big)\Pi^{\mu}\Pi^{\nu}\Pi^{\alpha}\Pi^{\beta}\pi^{(1)}_{\mu\nu}\sigma_{\alpha\beta}\bigg\}\bigg]~.
\label{phi2}
\ea
Putting (\ref{phi2}) into second order correction of shear viscous stress tensor ,
\be
\pi^{(2)\mu\nu}=g\int d\Gamma_p p^{\langle\mu}p^{\nu\rangle}f^{(0)}\phi^{(2)}~,
\label{shearstress2}
\ee
and combining with Eq.(\ref{shearstress1}) we have the evolution equation of shear stress tensor ($\pi^{\mu\nu}=\pi^{(1)\mu\nu}+\pi^{(2)\mu\nu}$) upto second order in velocity gradient,
\ba
\frac{\pi^{\mu\nu}}{\tau_{\pi}}=&2\beta_{\pi}\sigma^{\mu\nu}-\bigg[D\pi^{\langle\mu\nu\rangle}-2\pi_{\rho}^{\langle\mu}\omega^{\nu\rangle\rho}\nn
&+\lambda\pi_{\rho}^{\langle\mu}\sigma^{\nu\rangle\rho}+\frac{4}{3}\pi^{\mu\nu}(\partial\cdot u)\bigg]~,
\label{shearhydro1}
\ea
with $D\pi^{\langle\mu\nu\rangle}=\Delta^{\mu\nu}_{\alpha\beta}D\pi^{\alpha\beta}$ and $\beta_{\pi}=\frac{\eta}{\tau_{\pi}}$. 
The second order transport coefficients with $n^{th}$ power of momentum dependence in $\tau_R$ is given by,
\be
\tau_{\pi}=\tau_R^0\frac{(2n+4)!}{(n+4)!}=\frac{\eta/s}{T}\frac{5!(2n+4)!}{((n+4)!)^2}~,~~~~\lambda=\frac{2}{7}(2n+5)~.
\label{transcoeff1}
\ee
Eq.(\ref{phi2}) and (\ref{shearhydro1}) along with Eq.(\ref{transcoeff1}) are the main results of the current work.
It can be observed from Eq.(\ref{transcoeff1})
that $\tau_{\pi}$ which is defined as the relaxation time of shear viscous field and the microscopic relaxation time of particle distribution can only be
identical for momentum independent situation, i.e, with $n=0$. With increasing value of $n$, $\tau_{\pi}$ is becoming larger at a factorial rate making the
viscous field to decay at a slower rate with respect to the microscopic scale as expected. Where for linear ansatz $(n=0)$ we have $\tau_{\pi}=\tau_R^0$, for 
quadratic ansatz $(n=1)$ we have $\tau_{\pi}=6\tau_R^0$. The transport coefficients are, for linear ansatz $\tau_{\pi}=\frac{5\eta}{4P}$ and $\lambda=10/7$ 
as given in \cite{Jaiswal:2013npa} and with quadratic ansatz, $\tau_{\pi}=\frac{3\eta}{2P}$ and $\lambda=2$.

Next, let us explore the possibility of a fractional value of $n$ for the hydrodynamic evolution equation (\ref{shearhydro1}). The motivation comes from the work \cite{Dusling:2009df}
where in the context of radiative energy loss it has been shown that the momentum dependence of $\tau_R$ in defining the viscous correction to the phase-space 
distribution lies somewhere in between the two extreme limits of linear and quadratic ansatz assuming fractional powers as well. Since, all the transport
coefficients here are coming in terms of Gamma functions over $n$, then a fractional value of $n$ is theoretically acceptable as well. 
In \cite{Dusling:2009df} different dynamical theories have been discussed to have different fractional powers between these two extreme limits, where $n=1/2$
value corresponds to a two-flavor quark-gluon gas. Inspired by that fact, a fractional value 
$n=1/2$ has been taken here between the limiting values of linear and quadratic ansatz, for which $\tau_{\pi}=2.3\tau_R^0=\frac{1.31\eta}{P}$ and $\lambda=\frac{12}{7}$.

\subsection{Extended matching condition}

This so far straightforward formalism faces serious consequences while trying to pursue the Landau matching condition, $u_{\mu}u_{\nu}\Delta T^{\mu\nu}=0$, such that 
$u_{\mu}u_{\nu} T^{\mu\nu}=\epsilon$. This requires the integral $\int d\Gamma_p \tau_p^2 f^{(0)}\phi^{(r)}$ to be zero for every order $r$. From Eq.(\ref{phi1}) it is evident that
the first order correction to distribution function always satisfies matching condition irrespective of the value of $n$ (inner product property of irreducible tensors). However, 
it is the second order correction (\ref{phi2}) which behaves differently. Defining $\Lambda=u_{\mu}u_{\nu}\Delta T^{\mu\nu}$, and replacing (\ref{phi2}) in the expression of
$\Delta T^{\mu\nu}=g\int d\Gamma_p p^{\mu}p^{\nu}\delta f$, we find that,

\be
\Lambda=\frac{\tau_{\pi}}{2\eta}\bigg[(n+1)-\frac{((n+4)!)^2}{4!(2n+4)!}\bigg]\pi^{(1)\mu\nu}\pi^{(1)}_{\mu\nu}~. 
\label{elm}
\ee
It can be readily observed that $n=0$ is the only case where $\Lambda=0$ as also observed in \cite{Chattopadhyay:2014lya}. For momentum dependent $\tau_R$ ($n\neq 0$), $\Lambda$ 
clearly retains a non-zero value, which increases with increasing power of $n$. 
Before proceeding further to analyze the situation, we need to go through the origin and implications of the matching condition. The matching or fitting conditions
are actually constraints imposed on the dissipative part of energy-momentum tensor $T^{\mu\nu}$ and particle current $N^{\mu}$, in order to uniquely determine 
them from second law of thermodynamics. Violating matching conditions can result in thermodynamic instability (explained later) as well as altering thermodynamic equations
like (\ref{iden-1}) and (\ref{iden-2}). In rescue to this situation, a recently developed formulation of dissipative hydrodynamics by extending the matching conditions 
\cite{Osada:2011gx,Osada:2012yp} come in useful.

As nicely explained in the Appendix of \cite{Monnai:2009ad}, if the entropy 4-current $S^{\mu}$ has a term proportional to $\Pi u^{\mu}$ as non-equilibrium contribution 
($\Pi$ being bulk viscous pressure), the non-vanishing derivative $\partial(u_{\mu}S^{\mu})/\partial \Pi\mid_{\Pi=0}$ makes the system thermodynamically unstable. It 
means the system is not in a maximum-entropy configuration although equilibrium state has been used. However, as argued in \cite{Osada:2012yp}, a natural extension of equilibrium 
entropy current towards its off-equilibrium expression, may indeed contain such a term. Moreover, \cite{Monnai:2009ad} discusses, if the out of equilibrium distribution
function is taken as a combination of dissipative fluxes (ansatz for moment method), it is again natural that such a term to appear in $S^{\mu}$. Ref \cite{Osada:2012yp} treats the situation by nullifying this undesirable
contribution with the extended matching condition $u_{\mu}u_{\nu}\Delta T^{\mu\nu}=\Lambda$, retaining thermodynamical stability. In \cite{Osada:2012yp} the  
conformal contribution of $\Lambda$ turns out to be proportional to $\pi^{\mu\nu}\pi_{\mu\nu}$ with a proportionality constant $\tau_{\pi}/2\eta$, which is quite similar (apart 
from some numerical factors) to what has been obtained in Eq.(\ref{elm}).

Still two points are needed to be made here. Extended matching conditions turns the energy momentum tensor to take the following expression,

\be
T^{\mu\nu}=\big\{\epsilon u^{\mu}u^{\nu}-P\Delta^{\mu\nu}\big\}+\big\{\Lambda u^{\mu}u^{\nu}+\pi^{\mu\nu}\big\}~,
\ee
the first part being equilibrium and the later purely viscous correction.
It alters the energy density of the system by an additive factor of $\Lambda$ with respect to its equilibrium value. However, $\Lambda$ is a second order correction
and for a near equilibrium situation must have small contribution. It can be explained as a reflection of non-uniformity in the microscopic relaxation rate depending
upon the particle momenta. Greater the dependence of equilibrium restoration on particle momenta, larger the distortion in phase space distribution which finally results in 
generating macroscopic fields involving velocity gradients. Secondly, being a second order correction only, $\Lambda$ can not alter thermodynamic equations (\ref{iden-1}) and (\ref{iden-2}) within
second order dissipative theory, since the consequent corrections from $\Lambda$ will be at least of third order (correction in energy density equation (\ref{iden-1}) is $(D\Lambda+\Lambda\partial\cdot u)$
and in velocity equation (\ref{iden-2}) is $\Lambda Du^{\mu}$). However, for higher order theories the situation
can become non-trivial yet interesting.

It is noteworthy here that in a number of recent hydrodynamic analysis, this energy correction due to dissipative effects have been discussed 
giving the name of field redefinition, \cite{Bemfica:2017wps,Kovtun:2019hdm,Bemfica:2020zjp,Monnai:2018rgs} which state that conditions like Landau matching ($\Lambda=0$) is a mere choice of
frame for the relevant
hydrodynamic variables. However, this frame choice is inevitable since it defines the system's temperature or hydrodynamic velocity. \cite{Bemfica:2017wps} suggests to
resolve the ambiguity of defining them in presence of dissipation by considering $\epsilon$ (or $T$) and $u^{\mu}$ as auxiliary fields used to parameterize $T^{\mu\nu}$ which for small gradients represents the ideal values. Exact estimation for $\Lambda$ (the $n$ dependent
numerical factor) is somewhat arbitrary from these theories since it involves a range of frame choices permitted by well-posedness, causality, and stability
conditions \cite{Bemfica:2017wps}. In conformal limit with no conserved charges it is proportional to $\frac{\tau_{\pi}}{2\eta}\pi^{\mu\nu}\pi_{\mu\nu}$ at minimum order \cite{Bemfica:2020zjp}
as the same in the present work. Here the numerical factor turns out to be approximately $1$ (same as \cite{Osada:2012yp}) for quadratic ansatz ($n=1$). From $n=1$ to $0$ it smoothly reduces to zero which is 
Landau matching condition. So, this arbitrariness over $\Lambda$ has been somewhat constrained here using a momentum dependent microscopic theory.

\subsection{Equivalence between Chapman-Enskog and moment method}

Last part of the present work deals with attempting to establish an equivalence between this iterative technique involving gradient expansion of particle distribution
and the well known moment method. Following the formalism presented in \cite{Mitra}, the out of equilibrium part of the distribution function is expanded in particle
momentum basis which in the absence of bulk viscosity and conserved charges becomes $\phi=-C_{\mu\nu}\Pi^{\langle\mu}\Pi^{\nu\rangle}$. The unknown
coefficient $C^{\mu\nu}$ can be extracted by replacing $\phi$
in the expression of shear stress $\pi^{\mu\nu}=g\int d\Gamma_p p^{\langle\mu}p^{\nu\rangle}f^{(0)}\phi$ (note that this is the full expression for $\pi^{\mu\nu}$
unlike order by order expansion given by (\ref{shearstress1}) and (\ref{shearstress2})), which gives the total deviation $\phi=\frac{1}{8P}\pi_{\mu\nu}\Pi^{\langle\mu}\Pi^{\nu\rangle}$.
Next, we take the moment of relativistic transport equation (\ref{RTE}) by multiplying it with $g\tau_p^r\Pi^{\langle\alpha}\Pi^{\beta\rangle}$ and integrating over $d\Gamma_p$
($r$ is a power over the scaled energy which must be an integer). Now one thing is to note here. For $r=-1$, we get the following equation,
\ba
&&\int d\Gamma_p p^{\langle\alpha}p^{\beta\rangle}\dot{\delta f}=-\int d\Gamma_p p^{\langle\alpha}p^{\beta\rangle}\dot{f^{(0)}}\nn
&&-\int d\Gamma_p p^{\langle\alpha}p^{\beta\rangle}\frac{p\cdot \nabla f}{(p\cdot u)}+\int d\Gamma_p p^{\langle\alpha}p^{\beta\rangle}\frac{1}{p\cdot u}C[f]~.
\ea
This is same as Eq.(14) and (15) combined of \cite{Denicol:2014vaa} which boils down to the fact that $r=-1$ case represents the DNMR theory, where $r=0$ gives the 
usual Grad's 14-moment theory \cite{Monnai:2009ad}.

Performing the moment integrals (details in \cite{Mitra}), we obtain hydrodynamic evolution equation of shear viscous stress tensor in moment method as the following,

\ba
&&\frac{\pi^{\mu\nu}}{\tau_{\pi}}=2\beta_{\pi}\sigma^{\mu\nu}\nn
&&-\bigg[D\pi^{\langle\mu\nu\rangle}-2\pi_{\rho}^{\langle\mu}\omega^{\nu\rangle\rho}+\lambda\pi_{\rho}^{\langle\mu}\sigma^{\nu\rangle\rho}+\frac{4}{3}\pi^{\mu\nu}(\partial\cdot u)\bigg]~,
\label{shearhydro2}\\
&&\tau_{\pi}=\frac{\eta/s}{T}(r+6)~,~~~\lambda=\frac{2}{7}(2r+7)~,~~~\beta_{\pi}=\frac{\eta}{\tau_{\pi}}~.
\label{transcoeff2}
\ea
Eq.(\ref{shearhydro2}) is structurally exactly the same as (\ref{shearhydro1}) obtained from CE method. The corresponding transport coefficients (\ref{transcoeff1}) and (\ref{transcoeff2})
are observed to have identical expressions for two sets of choices of $n$ and $r$ values, $n=0, r=-1$ and $n=1, r=0$. Evidently, this is the reason why linear ansatz of iterative
gradient expansion \cite{Jaiswal:2013npa} has identical results with DNMR theory \cite{Denicol:2014vaa} and the results from quadratic ansatz of CE in Fig.(\ref{PLPT}) coincide with the
ultrarelativistic Grad's 14-moment results \cite{Mitra}. This equivalence is quite reassuring in acknowledging the fact that starting from the same microscopic theory (relativistic Boltzmann transport equation
in this case) different methods of obtaining hydro equations finally converge with each other. Any other correspondence between the two approaches (set of $n$ and $r$ values giving same result other 
than the mentioned)
is not known to the author.

\section{Results and discussion}

\begin{figure}[h]
\includegraphics[scale=0.36]{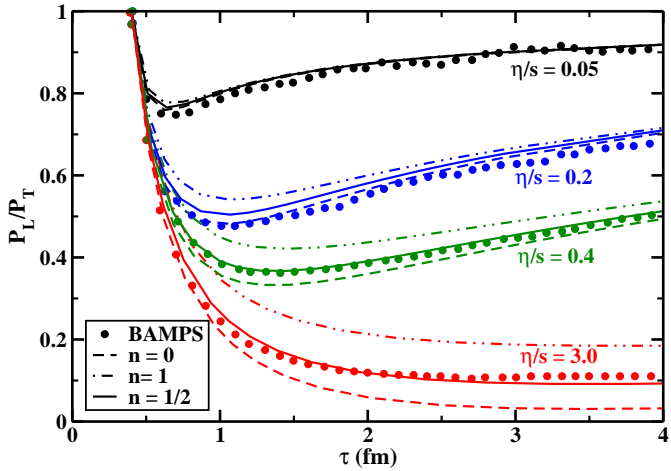} 
\caption{Time evolution of pressure anisotropy with different powers of momenta in $\tau_R$ at different $\eta/s$ ratio.}
\label{PLPT}
\end{figure} 

To have a quantitative idea how different powers of $n$ over particle momenta in $\tau_R$ affect the physical observables, Eq.(\ref{shearhydro1})
has been solved for a boost invariant Bjorken case for a massless Boltzmann gas with ultrarelativistic equation of state $(\epsilon=3P)$.
In terms of Milne coordinates $(\tau,x,y,\eta)$ with $\tau=\sqrt{t^2-z^2}$ and $\eta=tanh^{-1}(z/t)$ and considering the only independent
component of $\pi^{\mu\nu}$ to be $\pi^{\eta\eta}=-\frac{\pi}{\tau^2}$, we have the two following equations for energy density and shear pressure:
\ba
&&\frac{d\epsilon}{d\tau}=-\frac{\epsilon +P}{\tau}+\frac{\pi}{\tau}~,
\label{eqhydro1}\\
&&\frac{d\pi}{d\tau}=-\frac{\pi}{\tau_{\pi}}+\beta_{\pi}\frac{4}{3\tau}-\frac{(4+\lambda)}{3}\frac{\pi}{\tau}~.
\label{eqhydro2}
\ea
Eq.(\ref{eqhydro1}) and (\ref{eqhydro2}) have been solved with initial time and temperature at $\tau_i=0.4fm$ and $T_i=0.5 GeV$ with initial viscous pressure $\pi_i=0$,
for three values of momentum power $n$ in Eq.(\ref{relax}). Here, two limiting values of $n$ corresponding to linear $(n=0)$ and quadratic $(n=1)$ ansatz and one
in between fractional value $(n=1/2)$ has been considered with different $\eta/s$ values. The corresponding transport coefficients $\tau_{\pi}$ and $\lambda$ have
been obtained earlier for these three cases. Fig.(\ref{PLPT}) shows the proper time evolution of pressure anisotropy defined as $P_L/P_T=(P-\pi)/(P+\pi/2)$ for these
three values of $n$ and four sets of $\eta/s$ ratio. The obtained results have been compared with a numerical solution of the Boltzmann equation based on parton cascade
simulations (BAMPS) \cite{El:2009vj}, shown by solid circles. The dashed lines indicate the results from linear ansatz ($n=0$), the dot-dashed lines indicate the same
for quadratic ansatz ($n=1$) while the solid lines depict the $n=1/2$ case. The $n=0$ case as shown in \cite{Jaiswal:2013npa} under predicts the BAMPS data which
becomes prominent for large values of viscosity. $n=1$ case clearly over predicts the data a good deal showing even larger deviation from BAMPS for high $\eta/s$.
However, the $n=1/2$ situation remarkably agrees with BAMPS results even with large viscosity like $\eta/s=3.0$ throughout the evolution range. For small
viscosity like $\eta/s=0.2$ linear ansatz suffices to describe the dissipation, but with increasing viscous correction it is the fractional power of momentum dependence $n=1/2$
which provides a faithful representation of BAMPS data within the scope of second order dissipative hydrodynamic theory. 
It is worth mentioning here that in \cite{Dusling:2009df} the $n=1/2$ momentum dependence has been related to the dynamics of a two-flavoured quark-gluon gas
where the BAMPS data has been extracted for the same by a parton cascade model \cite{Xu:2004mz}.
This reasonable agreement of numerical data with fractional power of momentum dependence is very illuminating in the context of Ref \cite{Dusling:2009df} which argued that most of
interaction theories relevant for QGP lie between the two extreme limits of linear and quadratic ansatz and QCD kinetic theory predicts a momentum dependence within this 
range. So, momentum dependent relaxation time turns out to be a useful technique of kinetic theory which can capture the system dynamics reasonably accurately. 
It is to be noted here that in \cite{Jaiswal:2013vta,El:2009vj} the discrepancy between the hydrodynamics and kinetic theory results has been attempted to be resolved
using a third order hydrodynamic theory, but within the scope of second order relativistic hydrodynamics the current work is the best agreement known to the author.

\section{Conclusion and Outlook}

To summarize, a second order relativistic hydrodynamic theory has been developed with momentum dependent relaxation time approach using Chapman-Enskog formalism of
gradient expansion. The hydrodynamic evolution equation along with the transport coefficients have been estimated for two commonly used limiting cases - namely
linear and quadratic ansatz as well as for a fractional power of momentum dependence. The pressure anisotropy for the fractional power of momentum dependence shows an
impressive agreement with the numerical solution of Boltzmann equation indicating the system dynamics to lie somewhere in the middle of the two limiting ansatz. The anomaly
in Landau matching condition has been rescued with the help of an extended matching condition recently proposed. Finally, a correspondence between the iterative
technique of gradient expansion method and moment method has been established. 

A very interesting observation is revealed from this current analysis. From Eq.(\ref{shear1}) the $\eta/s$ can be written as, $(\eta/s)/(\eta/s)_0=(n+4)!/4!$
with $(\eta/s)_0=\tau_R^0T/5$ as the usual momentum-independent case ($n=0$), where the factor $(n+4)!/4!$ implements the modification due to momentum correction.
This variation in $\eta/s$ over $n$ values can be implemented in a macroscopic system by the temperature dependence of $\eta/s$ by using a temperature
dependent $n(T)$. This feature puts some constraints of $n$ values which is certainly desirable for the applicability of the theory. All the transport
coefficients expressed by Gamma functions over $n$, are fast growing for large values of $n$, so that any arbitrarily large value for $n$ will not be acceptable for the theory.
The $n$ value must lie in an acceptable range for which the transport coefficients follow the expected trend for the medium concerned.

\acknowledgments
The author thanks Subrata Pal for critical reading of the manuscript and helpful comments.

\end{document}